\documentstyle[color,aps,eqsecnum,psfig,graphicx]{revtex}

\newcommand{\eq}[1]{equation~(\ref{#1})}
\newcommand{\be}{\begin{equation}}
\newcommand{\ee}{\end{equation}}

\newcommand{\bea}{\begin{eqnarray}}
\newcommand{\eea}{\end{eqnarray}} 
\newcommand{\nonu}{\nonumber\\}
\def\eq#1{(\ref{#1})}
\def\hf{{1\over2}}
\def\ord{{\cal O}}
\def\la{\langle}
\def\ra{\rangle}

\def\vavr#1{\overline{#1}^V}

\def\addtext#1{{#1}}
\def\removetext#1{{}}

\def\at#1{\addtext{#1}}
\def\rt#1{\removetext{#1}}

\begin{document}
\title{Fate of the classical false vacuum}
\author{Sz. Bors\'anyi$^{a,1}$, A. Patk{\'o}s$^{a,b,2}$, J.
Polonyi$^{a,b,3}$, Zs. Sz{\'e}p$^{a,4}$}
\address{$^a$Department of Atomic Physics\\
E{\"o}tv{\"o}s University, Budapest, Hungary}
\address{$^b$Lab. de Physique Th{\'e}orique\\
Univ. L. Pasteur, Strasbourg, France}
\date{April 10, 2000}

\vfill
\footnotetext[1]{mazsx@cleopatra.elte.hu}
\footnotetext[2]{patkos@ludens.elte.hu}
\footnotetext[3]{polonyi@fresnel.u-strasbg.fr}
\footnotetext[4]{szepzs@cleopatra.elte.hu}
\maketitle
\begin{abstract}
Thermalisation of configurations with initial white noise power spectrum
is studied in numerical simulations of a \addtext{classical}
one-component $\Phi^4$ theory
in 2+1 dimensions, coupled to a small amplitude homogenous external field. 
The study is performed
for energy densities corresponding to the broken symmetry phase of the 
system in equilibrium. The effective equation of the order parameter motion
is reconstructed from its trajectory which starts from an initial value
near the metastable point and ends in the stable ground state. This
phenomenological theory quantitatively 
accounts for the decay of the false vacuum.
The large amplitude transition of the order parameter between the two minima
 displays characteristics reflecting dynamical
 aspects of the
 Maxwell construction.
\end{abstract}

\pacs{PACS: 11.10.Wx, 11.10.Hi}

\section{Introduction}
The reach of equilibrium from a metastable state covers a large
number of interesting effects from instabilities observed in
the mixed phase of first order phase transitions \cite{nucl}
to the inflation in the early Universe \cite{Boyan99}. The final state is 
reached in an irreversible process, the description of this
relaxation relies on out-of-equilibrium dynamics of field theories.

The decay of metastable states is usually discussed in the 
framework of the nucleation scenario \cite{Binder76}. It has been implemented 
in the form of saddle point expansions in classical \cite{Langer67}, and
quantum systems \cite{Coleman80}. This large amplitude instability is
\removetext{, however, only}
the first of the possible instabilities, suggested by mean 
field analysis, which \removetext{and}
consists of the creation of a bubble of the true vacuum
larger than the critical size, embedded into the false one. 

Another possibility, the instability against fluctuations with infinitesimal 
amplitude leads to the spinodal phase separation. 
A recent observation made it clear that soft fluctuations of these 
inhomogeneous unstable modes generate in equilibrium the Maxwell construction 
by their tree-level contribution to the renormalization group flow
\cite{Alex99}. The fluctuation induced flatness of the effective potential 
suggests the dominance of spinodal phase separation in equilibrium. 

Since the type of instability one observes, might depend essentially
on the time scale of the observation, a detailed investigation of the time 
evolution can separate the effects of the two kinds of 
instabilities. This is made possible by large scale computer simulations of 
the thermalisation process in closed systems. 

Another question, left open by the Maxwell construction in equilibrium
\cite{Callaway83}, 
concerns the structure of the vacuum with spontaneous symmetry breaking.
In fact, the effective potential is related to the probability
distribution of the order parameter and the Maxwell-cut applied to the former
suggests that the latter is also degenerate in the mixed phase.
Either we accept that the vacua with spontaneously broken symmetry 
correspond overwhelmingly to the mixed phase or 
a dynamical mechanism is seeked to eliminate the mixed phase from among the 
final states of the time evolution. 


In cosmology different slow-roll scenarios of inflation are being considered.
Recent studies of the dynamics of inflaton fields with large number of
components (large $N$ limit)
displayed for a first time a dynamical version of the Maxwell
construction \cite{Boyan99a}. 
The Hartree type solution of the quantum dynamical equations leads
to the conclusion that the order parameter might get rest with finite 
probability at any value
smaller than the position of the stable minimum of the tree level
effective potential, corresponding to the stabilisation of a mixed state.

Detailed investigations of the thermalisation phenomenon were performed
also in noisy-damped systems, coupled to an external heat bath
\cite{Alford93,Gleiser94a,Gleiser94b,Berera98}. The relaxation to thermal 
equilibrium of the space averaged scalar field (the order parameter) 
starting from metastable initial values has been thoroughly investigated. 
In these simulations the damping coefficient is treated as an 
external control parameter. Using the numerical solution of
the corresponding Langevin-equations the validity range of the analytical 
results for the homogenous nucleation mechanism has been explored. 

In this paper we  focus on an alternative description
of the decay process of the metastable vacuum state. The 
process is described exclusively in terms of the homogenous order parameter
(OP) mode. 
The evolution of the OP is studied in interaction with
the rest of the system as described by the reversible dynamical
equations of motion of the full system. 
Careful analysis of its dynamics allows us to explore the effects
of both kinds of the above mentioned instabilities. The transition
of the order parameter from the metastable to the stable vacuum is
induced by a homogenous external ``magnetic'' field, whose strength is
systematically reduced. No random noise is introduced to 
represent any external heat bath, the friction coefficient of the
effective order parameter dynamics is determined internally. 

Our results offer a "dualistic" resolution
of the competition between the nucleation and the spinodal phase separation 
mechanisms in establishing the true equilibrium. On the one hand,
we find that the statistical features of the decay of the
false vacuum agree with the results obtained
by expanding around the critical bubble. The microscopic mapping
of the field configurations during the relaxation supports the 
bubble creation scenario. Alternatively, the effective OP-theory
displays the presence of soft modes and produces
dynamically a Maxwell-cut when the time dependence of the transition
trajectory is described in the effective OP theory. 
We find that the larger is the system the smaller is the external field 
which is able to produce the instability. 
For infinite systems an infinitesimal field pushes the system
through the Maxwell-cut, where no force is experienced by the OP. Therefore
it will not stop before reaching the true homogenous ground state passing
by the mixed states with constant velocity. 
 
Our model, a spacelike lattice regulated \at{classical}
scalar $\Phi^4$ field theory 
in its broken symmetry phase is introduced in Section 2.
In Section 3 we describe the evolution of the system starting from
order parameter values near a metastable point which relaxes first to this 
state under the combined effect of parametric resonances and 
spinodal instabilities. The second stage of the
transition to the stable ground state is the actual focus of our paper.

Characteristic intervals of the observed order parameter trajectory 
are reinterpreted as being the solutions of some effective point-particle
equation of motion, which displays dissipation effects explicitly.
Our approach can be understood also as the real-time version of the lowest
mode approximation used for the estimation of finite size dependences in 
Euclidean field theory \cite{Zinn94,Binder84}. In this sense our approach 
can be considered also as the numerical implementation of a real time 
renormalisation group strategy. 

One of our principal goals is to reconstruct the
thermodynamics of the \at{classical}
"OP-ensemble" on the (meta)stable branches of the 
OP-trajectory (Section 4). Its dissipative dynamical equations near 
equilibria will be established.
On the transition trajectory we shall elaborate on the presence of the 
Maxwell construction in the effective dynamics describing the motion after 
nucleation (Section 5). The statistical
aspects of the approach to the equilibrium are established for 
reference and comparison in an Appendix. The conclusions of this investigation
are summarised in Section 6.

The results of this study can be usefully compared with classical 
investigations of metastability and nucleation in the kinetic Ising model
\cite{Binder74}. This system has first order dissipative dynamics 
by its definition. 
Still several relaxation features of the kinetic Ising model are comparable 
to our findings, since the ``numerical experiment'' performed in the two
models are essentially the same. 
Especially, the relaxation function of \cite{Binder74} 
is simply related to the order parameter we focus our attention.
In both cases in the mechanism of the bubble growth the aggregation of 
spontanously generated small size regions 
of the true ground state to its surface plays important role.

\section{Classical cut-off $\Phi^4$-theory on lattice}
The energy functional of a classical system in a two-dimensional box of size
$L_d$ coupled to an external magnetic field of strength $h_d$ is of the 
following form:
\be
E_d=\int d^2x_d\left[{1\over 2}\left({d\Phi_d\over dt_d}\right)^2+{1\over 2}
\bigl (\nabla_d\Phi_d\bigr )^2+{1\over 2}m^2\Phi_d^2+{1\over 24}\lambda\Phi_d^4
-h_d\Phi_d\right ].
\label{dimfulenergy}
\ee
The index $d$ is introduced to distinguish the dimensionfull quantities from
the dimensionless ones, defined by the relations (for $m^2<0$):
\bea
&
t=t_d|m|,\qquad x=x_d|m|,\nonumber\\
&
\Phi =\sqrt{\lambda\over 6}{1\over |m|}\Phi_d,\qquad h=\sqrt{\lambda\over 6}
{1\over |m|^3}h_d.
\eea
For the spatial discretisation one introduces a lattice of size
\be
L_d=Na_d=Na{1\over |m|}.
\ee
The energy functional of the lattice system can be written as
\bea
&
E\equiv {\lambda\over 6|m|^2}E_d={a^2\over a_t^2}\sum_{\bf n}\left [{1\over 2}
\left(\Phi_{\bf n}(t)-\Phi_{\bf n}(t-a_t)
\right)^2+{a_t^2\over 2a^2}\sum_{\bf i}\left(\Phi_{\bf n+\hat i}-\Phi_{\bf n}
\right)^2\right.\nonumber\\
&
\left.
-{a_t^2\over 2}\Phi_{\bf n}^2+{a_t^2\over 4}\Phi_{\bf n}^4-a_t^2h\Phi_{\bf n}
\right].
\label{dimlessenergy}
\eea
(Here we have introduced the dimensionless time-step $a_t$, which should be
chosen much smaller than $a$, and $\bf n$ denotes the lattice site vectors.) 
The equation of motion to be solved numerically is the following:
\bea
&
\Phi_{\bf n}(t+a_t)+\Phi_{\bf n}(t-a_t)-2\Phi_{\bf n}(t)-{a_t^2\over a^2}\sum_i
(\Phi_{\bf n+\hat i}(t)+\Phi_{\bf n-\hat i}(t)-2\Phi_{\bf n}(t))\nonumber\\
&
+a_t^2(-\Phi_{\bf n}+\Phi_{\bf n}^3-h)=0.
\label{latteq}
\eea

The initial conditions for Eq.(\ref{latteq}) were chosen as
\be
\dot\Phi_{\bf n}(t=0)=0,\qquad \Phi_{\bf n}(t=0)=\Phi_0+\xi\Phi_1.
\label{white}
\ee
The random variable $\xi$ is distributed evenly on the interval $(-1/2,1/2)$. 
Therefore the
starting OP-value is $\Phi_0$. The energy density $E/Na^2$ is controlled 
through the magnitude of $\Phi_1$. In this study we have chosen $\Phi_0=0.815$
and $\Phi_1=4/\sqrt{6}$. The latter corresponds to a temperature value 
$T_i=0.57$ in the
metastable equilibrium (from Eq.(\ref{tempeq})). This value is much below 
the critical temperature of the system ($T_c\simeq 1.5T_i$). 
It has been checked that at this energy density all other
choices of $\Phi_0>0$, for fixed $h$, find a unique metastable equilibrium. 

Eq.(\ref{latteq}) was solved with $a=1$ and with typical $a_t$ values in the
range (0.01-0.09). It has been checked that the statistical characteristics
of the time evolution is not sensitive to the variation of $a_t$, though the
``release'' time (the moment of the transition from metastability 
to the true ground state) 
in any single run with given initial conditions might change 
considerably under the variation of $a_t/a$. Three lattice sizes were 
systematically explored: $N=64,128,256$. Several single runs were performed
also for $N=512$ and $N=1000$ with the aim to analyze in more detail
some self-averaging
physical quantities on different portions of the trajectory under the
assumption of the ergodicity of the system. The magnetic field $h$ 
inducing the transition was
varied in the range $h\in -(0,0.08)/\sqrt{6}$. For the reconstruction of the
effective potential felt by the OP also positive values were 
chosen up to $h=0.5/\sqrt{6}$. The smaller the value of $|h|$ was fixed, 
the longer the ``release'' times have grown on the average. For this reason 
also the runs were prolongated with decreasing $|h|$, 
and for the smallest $|h|$ the length of a run reached up to 
$(10^6-10^7)|m|^{-1}$ until the transition took place. 

\rt{Table \ref{runstatistics} shows the transition event statistics for
each $(h,N)$ pair, used in this analysis. On smaller lattices, for a few
$h$ values very large number of transitions were observed, in order to
clarify the nature of the corrections to the nucleation mechanism.}
\at{For a careful comparison of our transition statistics with the
generally used statistical nucleation theory, and also for
understanding the systematics of its change when $h$ has been
diminished a large number of ($h$, $N$) pairs were used in this
analysis. Altogether \hbox{24~422} transition events have been
recorded (for \hbox{$N=64$:~~16~908}, \hbox{$N=128$:~~2~903},
\hbox{$N=256$:~~4~411}).}
For the largest systems at the smallest $h$ the event rate was 
1-2/day/ 400 MHz-processor.
 

\section{Time-history of the order parameter}
\begin{figure}                                       
\includegraphics[width=8.5cm]{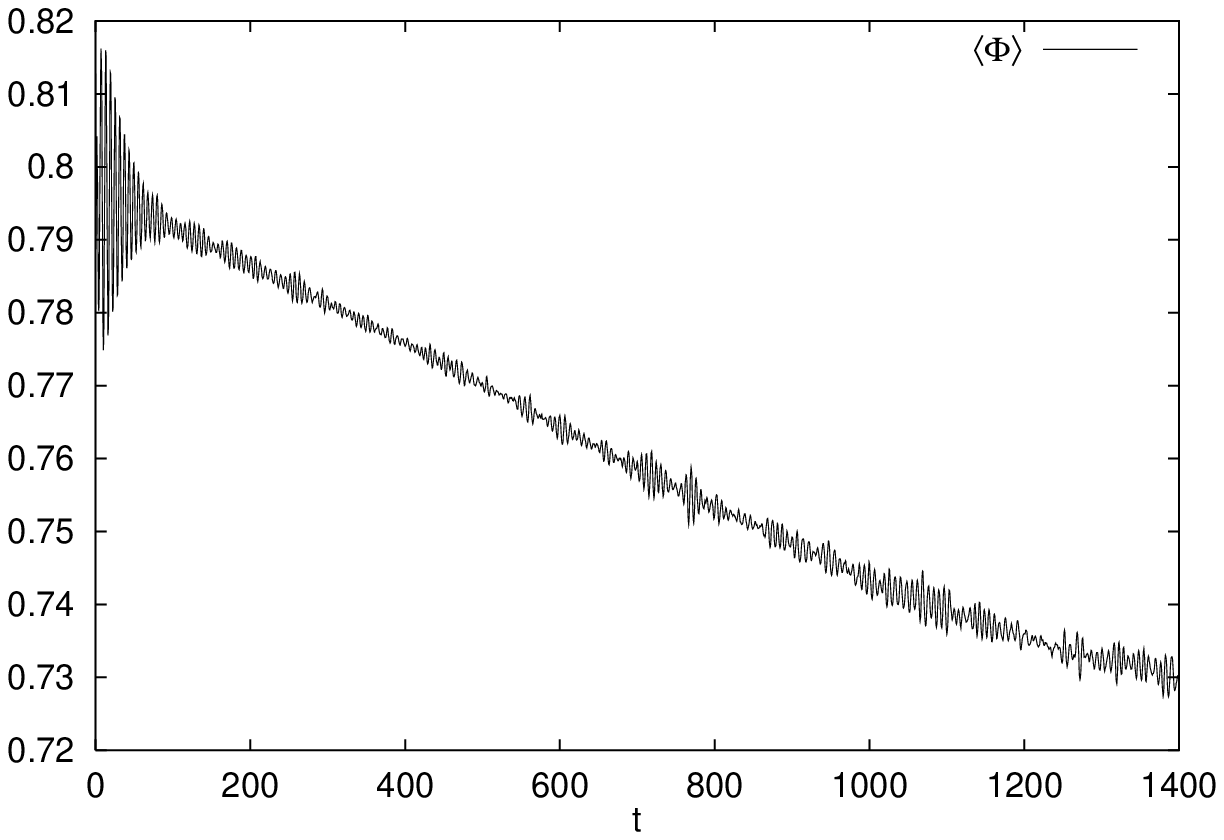}
\includegraphics[width=8.5cm]{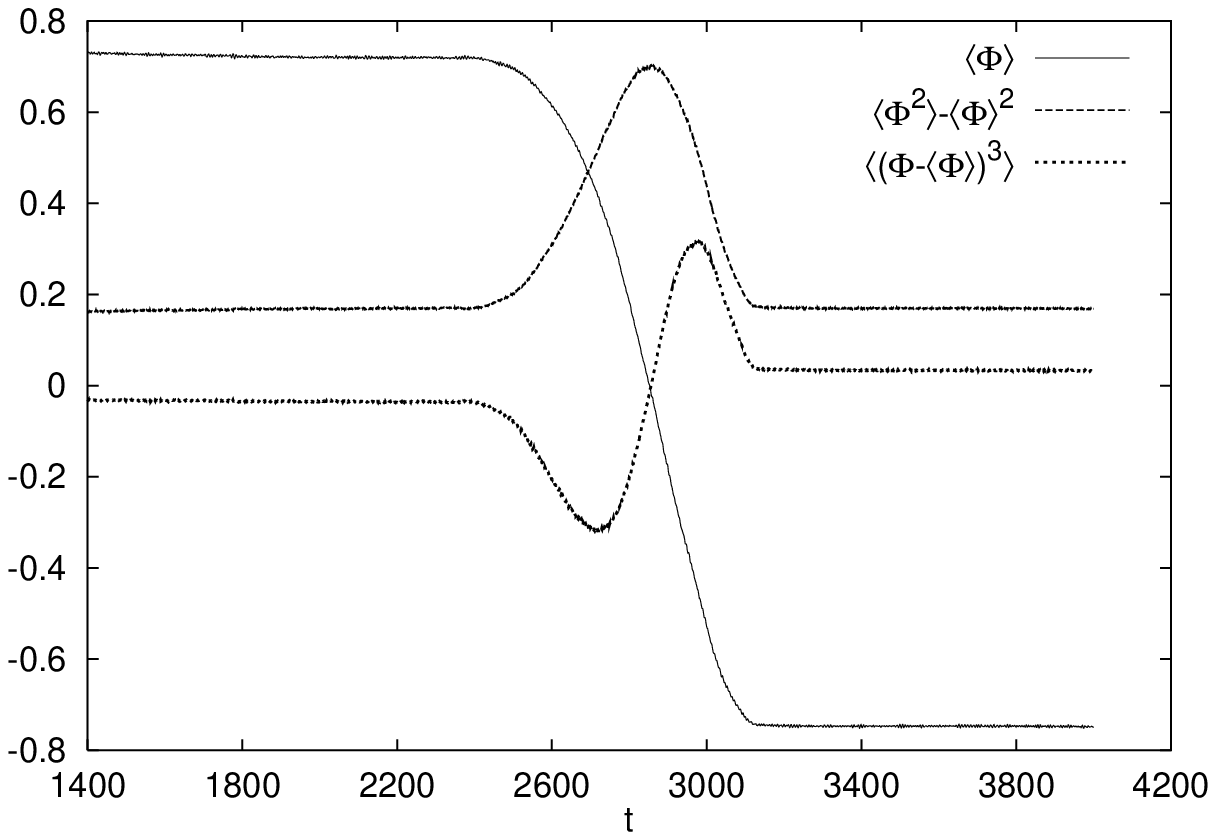}
\caption{\label{time_hist}The time evolution of the order parameter,
its MS fluctuation and the third moment.
The example is selected from runs on a $N=512$ 
lattice with $h=-0.04/\sqrt{6}$ external source strength.}
\end{figure}

\begin{figure}
\begin{center}
\includegraphics[width=8.5cm]{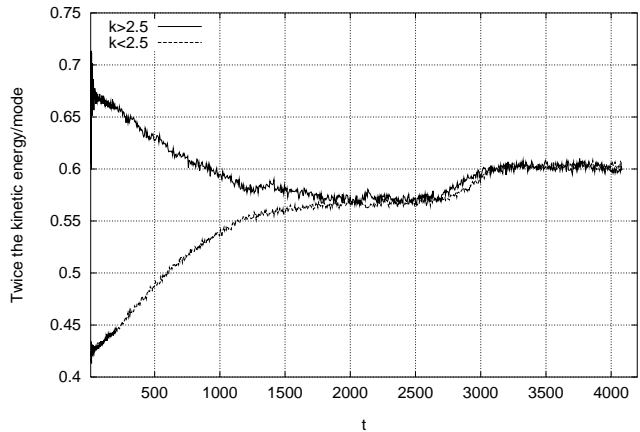}
\end{center}
\caption{\label{k_hist}The time evolution of the kinetic energy content of
the $|{\bf k}|>2.5$ and $|{\bf k}|<2.5$ regions averaged over the
corresponding $|\bf{k}|$-intervals.
The example is the same as in Fig. 1.}
\end{figure}

A typical OP-history is displayed in Fig.\ref{time_hist}. 
In the same figure we show also the history of the OP mean square 
(MS)-fluctuation ($\langle\Phi^2\rangle -\langle\Phi\rangle^2$) and
of its third moment ($\langle (\Phi -\langle\Phi\rangle )^3\rangle $). 
The evolution
of the non-zero $\bf k$ modes is demonstrated in Fig.\ref{k_hist}, where
the averaged kinetic energy content of the $|{\bf k}|<2.5$ and $|{\bf k}|>2.5$
regions is followed. Although the separation value is somewhat arbitrary, 
namely it divides into two nearly equal groups the spatial frequencies
available in the lattice system,
the figure demonstrates the most important features of the evolution of the 
power in the low-$|{\bf k}|$  and high-$|{\bf k}|$ modes.

In general, five qualitatively distinct parts of the trajectory can be 
distinguished, although some of the first three might be missing for 
some initial configurations and/or magnetic field strengths.

The OP-motion usually starts with large amplitude damped oscillations.
The ``white noise'' initial condition of Eq.(\ref{white}) corresponds to a 
$\bf k$-independent Fourier amplitude distribution, therefore the initial 
distribution of the kinetic energy is $\sim\omega^2(|{\bf k}|)$.
During this period, in the power spectrum of the kinetic energy, first 
a single sharp peak shows up at a resonating $|{\bf k}|$-value ($|{\bf k}|
\sim 1.5$), which breaks up into several peaks 
($|{\bf k}|<1$) at later times due to the non-linear interaction of the modes. 
At the end of the first period the whole $|{\bf k}|<1$ range 
gets increased power, the $|{\bf k}|>1.5$ part of the power spectrum does
not seem to change. 

Next a slow, almost linear (modulated) decrease of the OP follows. 
At the same time its MS fluctuation increases linearly. 
In Fig.\ref{k_hist} an energy flow towards the low-$|{\bf k}|$ part can be 
observed, which proceeds through the excitation of single modes in this 
part. As a result the minimum of the effective potential is continously 
shifted to smaller $\Phi$-values, as if the temperature would gradually 
increase (see the left part of 
Fig.\ref{time_hist}). The full (microscopic) kinetic energy density shows in 
this period less than 5\% variation. In view of the picture based on 
the Maxwell construction it is rather surprising
that independently of the initial conditions the OP converges towards
a well-defined absolute value, depending only on the total energy density.

On the third portion the average value 
of the OP and its moments stay constant
(see the right picture in Fig.\ref{time_hist}). The average energy content
of the low and high-$|{\bf k}|$ part of the spectra is nearly the same. 
This suggests the establishment of a sort of thermal (meta)equilibrium. 

In terms of the terminology introduced for the inflation, 
the first period leading to this
(quasi) -stationary state can be called {\it preheating}, and the second 
{\it reheating}.  Directly before the 
moment of the transition to the true vacuum
a peak appears in the power spectrum in the narrow neighbourhood of ${\bf k}=0$
with varying position in time. On the snapshots of the real space 
configurations a set of randomly distributed small bubbles of the true
ground state appear with a radius increasing in time, until one of the 
bubbles exceeds the critical size.

The fourth portion of the motion is the transition itself. The value 
of the OP MS-fluctuation increases by about a factor of three and 
the temporal width of 
this transient increase measures very well the transition time. The
transition time decreases on larger lattices, the height of the
jump in the OP-fluctuation is not sensitive to the lattice size. 
Also the third reduced moment shows a characteristic variation. 
An increase of the temperature proceeds smoothly
during the transition of the OP to its stable value (Fig.\ref{k_hist}).
The slight separation of the two curves in Fig.\ref{k_hist} gives a 
feeling on the degree of uniformity of the temperature variation of the 
different modes.

The last portion of the trajectory represents stable (thermalised) 
oscillations around the true ground state. 
Here a complete equilibration of the power spectrum can be observed
(see Fig.\ref{k_hist}) corresponding to a somewhat increased temperature
$(T\sim 0.6)$.

\section{Motion near the (meta)\-stable point}
Our analysis of the motion around the (meta)stable value of the order 
parameter explores the consequences of assuming the ergodicity hypothesis for
\at{a sufficiently long}
finite time interval,  \at{after the system has already reached the
equilibrium}. \at{
The equilibrium is characterised by a limiting probability density in
the configuration space. The averaging with this density should provide
the same value as the one yielded by a single long time evolution when 
subsequent configurations are used in constructing the
statistics. Our ``statistical system'' is now a single degree of
freedom, the order parameter of the lattice system, interacting with
all other ($k\ne0$) modes.} 

\at{
The effective equation of motion can be thought to result from
the application of a ``molecular dynamical renormalization group'', to our
microscopical equations. 
The blocking in space is performed by projecting the field configuration
$\Phi({\bf x},t)$ on the OP $\Phi (t)$. It represents the infrared (IR) end 
point of such a blocking whose effective theory
is now reconstructed from the actual time dependence found numerically.
Combining ergodicity of the full system with the renormalization group (RG)
concept we arrive to the conclusion that ensemble averages
of any OP-function coincide with time averages of the same function. 
} 
\rt{
In case of the motion around metastable minima we attempt
to detect signatures of the 
metastability in the period directly preceding the start of the transition of 
the order parameter to the stable position.}

\at{
In view of the RG concept we look for an effective equation of
motion where the value of OP is determined exclusively by its values preceding
in time. Assuming the absence of long memory effects a ``gradient''
expansion in time can be envisaged, leading to a local differential
equation.}
\at{
We introduce into this
phenomenological ``Newton-type'' equation a term violating time-reversal 
invariance. We are not able to derive it from the original system, we only 
wish to test its presence. The sustained motion of OP, however, requires, 
in this case the presence of a random ``force'' term, too. 
The deterministic part of the ``force'' is expected to be
related to the equilibrium effective potential, since this object determines
the stationary probability distribution for the OP near equilibrium.}

\at{The above considerations lead us to write down the  equation of
motion, which represents a linear relation between the acceleration and the
velocity of the order parameter:
}
\be
\ddot{\Phi}_d+\eta_d(\Phi_d)\dot{\Phi}_d-h_d
+{dV_{\textrm{eff}}(\Phi_d)\over d\Phi_d}=\zeta_d,
\label{eqfriction}
\ee
where $\zeta_d$ is a noise term. In the corresponding dimensionless equation
of motion the following new rescaled quantities will appear:
\be
\eta_d=|m|\eta, \qquad \zeta_d=\zeta |m|^3\sqrt{6\over\lambda}.
\ee
\at{
The fitting procedure for the coefficients on the left hand side of
Eq.~(\ref{eqfriction}) was the following.
For a given interval of time $I_t$ the region of the order parameter space
visited by the system $\left\{\Phi(t)|t\in I_t\right\}$ was divided into small 
bins. Having defined the time set 
$T_{\Phi_b}=\left\{t\in I_t|\Phi_b<\Phi(t)<\Phi_b+\Delta\Phi\right\}$ 
corresponding to a given bin, the linear relation
$\ddot\Phi(t_b)=-\eta(\Phi_b)\dot\Phi(t_b)-f(\Phi_b)$
was fitted with the method of least squares
using the $\dot\Phi(t_b)$, $\ddot\Phi(t_b)$ data measured at time moments 
$t_b$ belonging to $T_{\Phi_b}$. We have obtained in this way the coefficient 
functions $\eta(\Phi_b), f(\Phi_b)$. Once the functions $\eta(\Phi)$ and
$f(\Phi)\equiv-h+V_{\textrm{eff}}'(\Phi )$ are determined, we evaluate 
for each time $t$ the expression 
$\ddot\Phi(t)+\eta(\Phi(t))\dot\Phi(t)+f(\Phi(t))$. Its actual value
determines the random noise function $\zeta(t)$, whose statistical
features (autocorrelation) should be extracted from the data.}
\rt{
The region of the order parameter space 
visited by the system was  divided into small bins and Eq. (\ref{eqfriction})
was fitted within each of them without the noise term. Once the functions
$\eta (\Phi )$ and $f(\Phi )\equiv-h+V_{\textrm{eff}}'(\Phi )$ are determined, 
the moments of the noise
can be identified with the corresponding moments of the deviation at a given 
time of the left hand side of the
equation of motion (\ref{eqfriction}), evaluated with the fitted
$\eta (\Phi )$ and $f(\Phi )$, from zero.}

The {\it effective force} $f(\Phi )$ calculated from the time-average of the
oscillatory motion around the equilibrium, is expected to
agree with the force coming from the theoretically determined
finite temperature effective potential calculated perturbatively
in the cut-off two-dimensional field theory for some appropriately chosen
value of the temperature \cite{Kapusta89}. With one-loop accuracy the
expected equality reads:
\be
f(\Phi )_{measured}=-h+{d\over d\Phi}\left[V(\Phi )+{T\over 8\pi}V^{''}
(\Phi )\left(1+\log{\Lambda^2\over V^{''}(\Phi )}\right)\right]_{theory}
+{\cal O}(T^2).
\label{fforce}
\ee
The expressions on the right hand side of this equation are connected
to the dimensionfull quantities of the original one-loop computation in the
following way:
\be
T={\lambda\over 6|m|^2}T_d,\quad 
V(\Phi )=-{1\over 2}\Phi^2+{1\over 4}\Phi^4={\lambda\over 6|m|^2}V_d,
\quad \Lambda ={\pi\over a}.
\ee
The dimensionless temperature is defined by the time-average of the 
kinetic energy based on the assumption that in the effective theory of the
OP it has the usual expression in terms of $\dot\Phi (t)$:
\be
\lim_{t\rightarrow\infty}{1\over t}\int_0^tdt'{1\over 2}
(\dot\Phi (t',x))^2\equiv{1\over 2}\overline{\dot\Phi^2}^{t}={T\over 2}.
\label{tempeq}
\ee
One should note that this definition in the dimensionfull
version implies a ``Boltzmann-constant'', $|m|^2$ multiplying $T_d$.

By measuring the order parameter average for different values of the
external field $h$ we obtain the magnetisation curve of the system.
In the numerical work we restarted the computation at different values
of the external magnetic field but one might as well change $h$ adiabatically
and measure the force law at each (quasi)equilibrium point. The
results should agree in the stable regime, up to the phenomenon of the
hysteresis.
The resulting curve can be viewed as the numerical Legendre transformation
by identifying the external source $h$ with the derivative of the
effective potential,
\be
-h+V'_{\textrm{eff}}(\la\Phi\ra_{h,measured})=0.
\label{equilib}
\ee
It was found that $f(\Phi )_{measured}$ is always vanishing at 
$\Phi=\la\Phi\ra_{h,measured}$,
thus the force acting on the order parameter at the
equilibrium position $\la\Phi\ra_h$ induced by the external 
source $h$ is indeed always $-V'_{eff}(\la\Phi\ra_h)$. 

\begin{figure}
\begin{center}
\includegraphics[width=12cm]{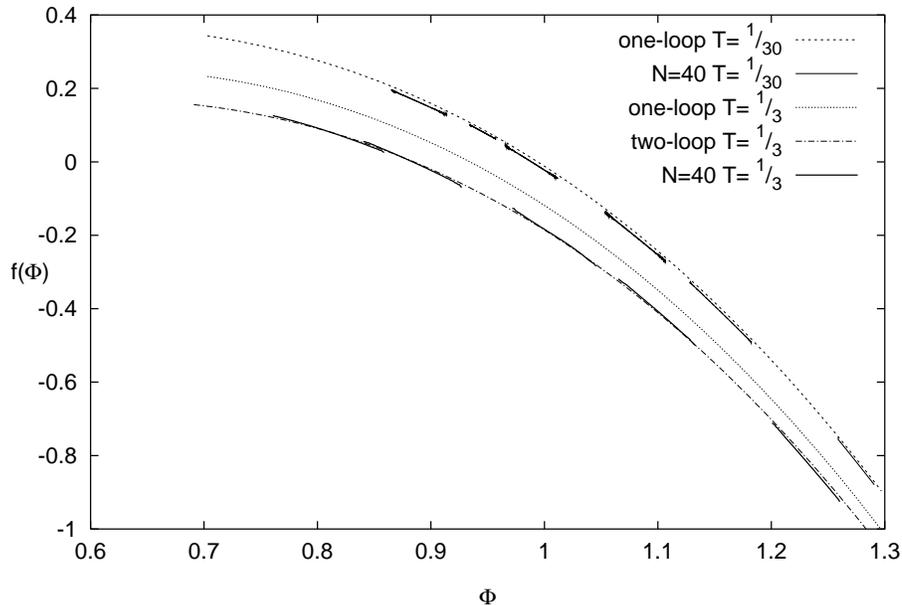}
\end{center}
\caption{\label{fphi} The force $f(\Phi)$ as the function of the order
parameter on $40\times40$ lattices with
 $T=1/30$ and $1/3$. The force is measured by shifting the center of motion
to different $\Phi$ values by applying appropriate $h$ fields to the system.
The finite size effects are negligible, see 
text. For comparison we display the force arising from the derivative 
of the one-loop and for $T=1/3$ also of
the two-loop expressions of the effective potential.}
\end{figure}

The two sides of the relation (\ref{fforce}) are shown in Fig.\ref{fphi}.
To the values of the external field used in preparing Fig.\ref{fphi}
($h\sqrt{6}=-0.5, -0.02, 0, 0.5, 1, 2)$ 
single runs were selected by the requirement
that the system stayed in the metastable vacuum up to the time $10^6$.
The force shown in Fig.\ref{fphi} demonstrates that not only the
equilibrium positions but also the fluctuations of the OP
are governed by the {\it static} effective potential. Similar measurements,
performed on $100\times 100$ lattices showed no finite size effects
in the stable regime, $h>0$. The error of
$f(\Phi )$ is not shown in the Figure, since the typical values
($\sim 0.004$ for $T=1/3$ and $\sim 0.002$ for $T=1/30$) are too small to be
displayed.

Another piece in the effective equation of motion \eq{eqfriction}
is the {\it friction term} whose presence indicates the dynamical
breakdown of the time inversion symmetry. The friction coefficient
$\eta(\Phi)$ proves clearly non-vanishing and shows only weak 
dependence on the actual value of the OP around its equilibrium position. 
The breakdown of the time-reflection symmetry in a closed system must arise 
only in presence of
infinitely many degrees of freedom. Till then only statements on the 
Poincar{\'e}-time can be made. Thus our non-zero results for $\eta$
require further clarifications. 

The point is that there are two types of
infinities, controlled by the {\it temporal} UV and the IR cutoffs, 
respectively. The spontaneous symmetry
breaking is driven by the IR modes, and the non-trivial minima of the
potential energy arise from the presence of infinitely many degrees of 
freedom in the IR (thermodynamical limit). On the contrary, 
the dynamical symmetry breaking
\cite{dynam} is the result of the effects of the derivative terms
in the action and the infinitely many UV modes (continuum limit). 
The breakdown
of the time inversion, being related to a time-derivative term in the
effective equation of motion, should come from the UV, the short
time behaviour of the system. 

In fact, one expects no friction when the UV cutoff, $a_t$ is \at{so} small,
\at{that not enough energy can be dissipated during such a short time. In
quantitative terms one should have}
\be\label{tau}
a_t>\tau={2\pi\over\sqrt{{8\over a^2}+2}}+\ord{(T)}, 
\ee
where $\tau$ is the time scale of the fastest mode in the system. 
The right hand side relation of Eq.\eq{tau} gives \at{an estimate of} the 
maximal frequency from the free dispersion relation 
$p_0^2=4(\sin^2p_xa/2+\sin^2p_ya/2)/a^2+M^2(T)$
of the lattice hamiltonian system for fixed $a$ in the limit $a_t\to0$.
The fast modes absorb energy from the OP
in a single time step for $a_t>\tau$, and the friction term should appear 
in the effective equation of motion.

\begin{figure}
\begin{center}
\includegraphics[width=12cm]{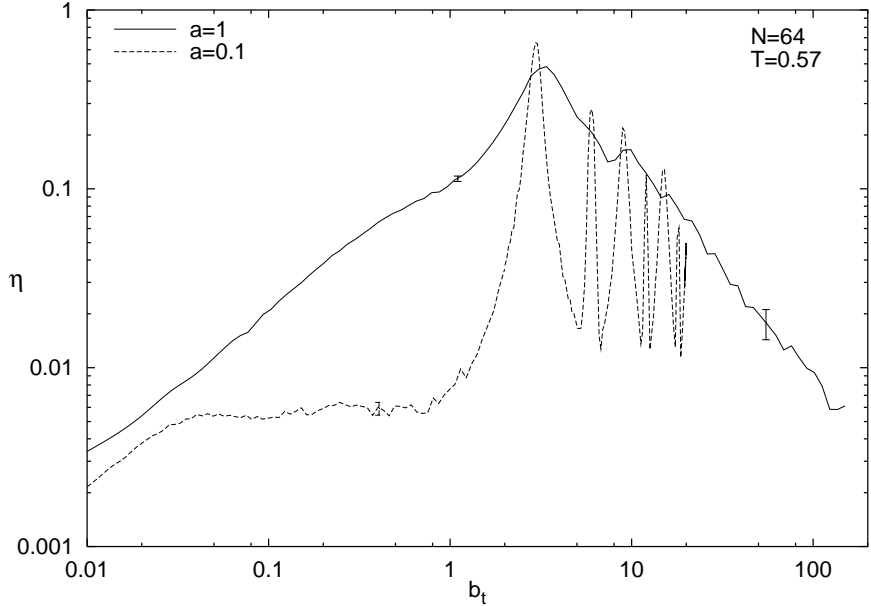}
\end{center}
\caption{\label{atdep}
The friction $\eta (\la\Phi\ra_{h=0})$ as the
function of the timelike lattice spacing $b_t=na_t$. The time scale of the
fastest
mode is $\tau\approx2$ and $0.22$ for $a=1$ and 0.1, respectively,
according to the perturbation expansion. Typical error bars
appear on the curves both for the IR and UV regimes.}
\end{figure}   

\at{With help of the temporal blocking} 
\be\label{tblo}
\Phi(t)\to\Phi_{n}(t)={1\over n}\sum_{k=-n/2}^{n/2-1}\Phi(t+ka_t)
\ee
\at{one can construct the trajectories $\Phi_{n}(t)$
corresponding to larger values of the time
cut-off, $b_t=na_t$. Such blocking}
was performed in time up to $n=2000$ and a discrete time
 equation of motion of the form
Eq.(\ref{eqfriction}) was reconstructed for the blocked trajectory
\at{$\Phi_n(t)$}. 
 The non-trivial dependence of $\eta$ on the temporal cutoff $a_t$ is shown
in Fig. \ref{atdep}. 

One can distinguish two regimes
separated by a crossover apparently independent of
 the lattice size, located at $b_t\approx0.2\tau$. A scaling behavior
is observed on the UV side, where 
$\eta(b_t)$ tends to zero with a critical exponent close to 1 and the force
being constant. This result should be independent
of the actual blocking details. In the other regime, on the
IR side $\eta$ goes over first
into a $b_t$-independent regime, corresponding to the saturation 
of the energy transfer from the OP and giving a stable, microscopic 
definition of the friction coefficient. Finally, in the far IR part a 
qualitatively different oscillatory behaviour sets in. 

The location of the crossover from the UV scaling regime to the plateau can
be understood by writing the fluctuation dissipation theorem in the
corresponding discretised form: $\langle\zeta^2\rangle =2\eta T/b_t$.
The linearly increasing regime of
$\eta (b_t)$ implies constant second moment for the noise. 
When $\eta (b_t)$ reaches the plateau
the second moment of the noise decreases like $1/b_t$. The crossover 
therefore is located at the autocorrelation time scale of the noise.

A qualitative interpretation of the oscillatory IR regime can be based on the 
observed
small amplitude beating phenomenon in the OP trajectory. This can be recognized
by closer naked eye inspection of the left side of Fig.\ref{time_hist}, which
persists further also on the right side. It is reflected in the OP 
autocorrelation function, too, since in the course of the blocking 
an interference effect occurs on the right hand side of Eq.\eq{tblo} 
due to this regularity. This feature is relevant to the value 
of $\eta (b_t)$, responsible for the decay of all kinds of fluctuations. 

The appareance of peaks in $\eta(b_t)$ at both the
maximal destructive and constructive interferences can be modeled
semi-quantitatively by
identifying the  beating part of the OP-motion 
${\textrm{Re}}\delta\Phi={\textrm {Re}}\delta\Phi_0\exp(i\omega t)$
with the stationary solution of a single weakly damped
driven harmonic oscillator,
$\ddot{\delta\Phi}+\eta\dot{\delta\Phi}+\omega_0^2\delta\Phi=
f_0\exp(i\omega t)$.
The blocking acts on the trajectory $\delta\Phi(t)$
as $\delta\Phi(t)\to u\delta\Phi(t)$, where 
$u=(\exp(i\omega b_t)-1)/i\omega b_t$.
It does not change the relative phase of the driving force and 
of the blocked oscillation amplitude, leading to a relation
between the parameters of the original and the blocked equation of motion:
\be
-\omega^2+i\eta\omega+\omega_0^2=-\omega^2u^2+i\bar\eta\omega u+\bar\omega_0^2.
\ee

The friction coefficient for the blocked trajectory turns out to be
$\bar\eta=(\eta+\omega \textrm{Im}(u^2))/\textrm{Re}(u)$. It is easy to see
that $\textrm{Re}(u)$
is vanishing at maximal constructive and destructive interferences.
This provides singularities in $\bar\eta$. 
Whenever $\textrm{Re}(u)=0$ we have $\textrm{Im}(u^2)=0$
and numerator changes sign in the vicinity of the singularity. Thus
$\bar\eta>0$ apart for a short time interval around the singularities where
the non-harmonic features should stabilise the fluctuations and
keep $\bar\eta>0$, as observed in our simulation. 

The $a$ dependence appearing in
Fig.\ref{atdep} arises from the following two effects. 
One is that for larger $a$ 
the maximal oscillation frequency is smaller and the time resolution 
of the system becomes cruder. Another is that larger
$a$ represent bigger physical volume, many more 
soft modes and less harmonic system, 
which tends to invalidate the simple picture based on a single 
harmonic mode. As a result the effects of oscillatory nature will be 
smeared, as one clearly recognizes in the figure.

The quality of any proposed deterministic equation of motion (e.g.
equations similar to Eq.(\ref{eqfriction}) with zero on the right hand side),
can be judged by the amplitude and the autocorrelation of its error term,
{\it the noise term} $\zeta$. The amplitude of the noise was found
at least two-three order of magnitude below the average level of the force 
as fitted to Eq. \eq{eqfriction}. 
The autocorrelation
function of the noise of Eq. \eq{eqfriction} appeared to be local, 
approximately of the form $\la\zeta_d(t)\zeta_d(t')\ra\approx\delta''(t-t')$. 
The status of the fluctuation-dissipation theorem will be investigated for 
more complex field theoretical systems in future investigations.

\begin{figure}
\begin{center}
\includegraphics[width=12cm]{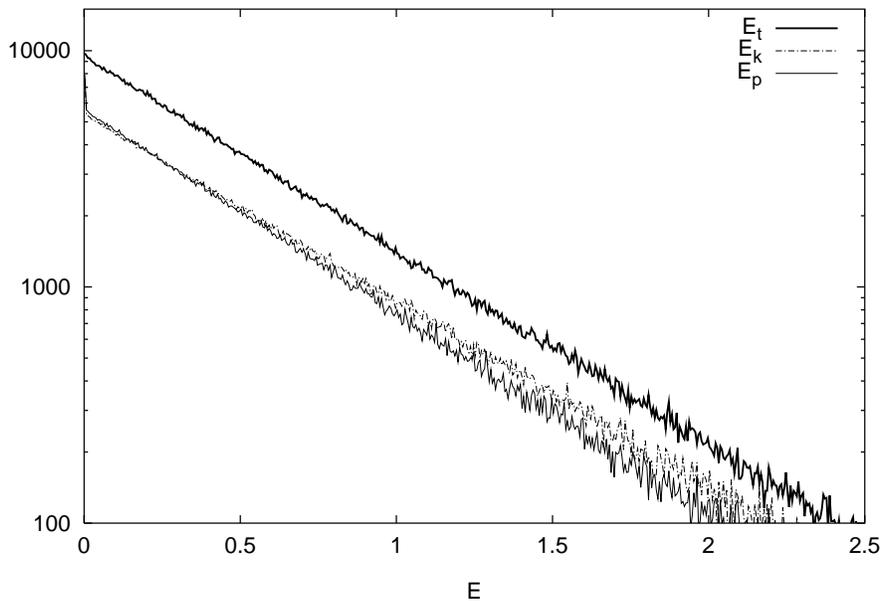}
\end{center}
\caption{\label{th} The histogram of the energies $E_t$, $E_k$
and $E_p$ of Eq.(\eq{ens}) on a $256\times256$ lattice with $h=2/\sqrt{6}$.}
\end{figure}

The selfconsistency of the definition of the temperature in Eq.(\ref{tempeq})
and also the establishment of the thermal equilibrium
can be tested further by plotting histograms for the following quantities:
\bea\label{ens}
E_k&=&\hf\left({d\Phi\over dt}\right)^2,\nonu
E_p&=&{\mu^2\over2}(\Phi-\langle\Phi\rangle_h)^2,\\
E_t&=&E_k+E_p,\nonumber
\eea
where $\mu^2$ is the slope of the force as the function of the order 
parameter, determined numerically. 
Typical results for the energy-histograms 
are shown in Fig.\ref{th}. 
It shows perfect agreement of all slopes and good agreement with the 
expectations based on equipartition of the energy between the kinetic and
the potential parts. We have checked for several temperatures, that the 
temperature
determined by these histograms and from the $|{\bf k}|$-spectra of the kinetic 
energy agree. This piece of information is parallel to the recent detailed
investigations of the thermalisation in $(1+1)$-dimensional $\Phi^4$-theories
\cite{parisi97,aarts00}.

It is worthwhile noting that the relaxation time from
a given initial condition to the thermally distributed state like in this
figure takes at least one order of magnitude longer time for
systems in the symmetric phase compared to the spontaneously
broken case. At higher energy density (temperature)
one would expect larger collision frequency, therefore shorter
thermalisation time. The opposite result hints to the importance of slow, soft
modes, whose presence is 
due to the symmetry breaking mechanism. We shall argue in the
next Section that these modes are responsible for the realization of the
Maxwell-cut in the potential term of the effective equation of motion
for the OP. What we find remarkable is that these modes are present 
not only in the mixed phase but also near (meta)stable equilibria, 
among the dynamical 
fluctuations around the ordered vacuum in contrast to the massive perturbative 
excitation spectrum in the equilibrium.

We complete the analysis of this section with the remark, 
that after the transition to the stable vacuum shortly a 
thermalised distribution is recovered for the OP at somewhat increased
temperature which agrees with the value given by the equipartition.

\section{Jump from the metastable to the stable vacuum}
As it has been emphasized in the Introduction, there are (at least)
two different descriptions of the transition from the
metastable state to the stable one. \at{In the first approach an}\rt{The}
expansion around the
lower pass, the \rt{``sphaleron''}\at{critical bubble (bounce)}
configuration yields a detailed
space-time picture and the transition rate by means of the 
analytic continuation of the potential experienced by the OP near the
(meta)stable position \cite{Langer67,Coleman80}.

Another possibility is to provide for the OP a probabilistic
description, obtained by the elimination of all other degrees of freedom
in some kind of blocking procedure. 
This description is based usually on some underlying Master equation for the
probability distribution of the OP and the
resulting Fokker-Planck equation \cite{vankampen}. The transition
to the stable state appears in this approach formally as a tunneling solution
of the Fokker-Planck equation. The probabilistic feature
of the dynamics of the OP is supposed to arise from assuming \at{a
statistical ensemble of} \rt{averaging over} the initial conditions. 

In our simulation we find results analogous with the predictions
of the probabilistic description by analysing the OP-motion starting from
a single, well defined initial condition. 
One might wonder at this point if it is possible to 
understand the probabilistic tunneling of the OP by following
the system from a unique initial condition. The self averaging in
time can not be used for this argument since such a transition
occurs only once during the evolution in a (quasi)irreversible manner.
\at{We have to develop a third approach to the \hbox{metastable$\to$stable}
transition.}

In general, the effective equation of motion for $\Phi(t)$ reflects the typical
landscape of the microscopic potential energy functional around the
actual point $\Phi({\bf x},t)$ in the configuration space. 
The classical origin of what appears as a tunneling on the Fokker-Planck
level must be the arrival of $\Phi({\bf x},t)$ to the vicinity of some
narrow valley opening up towards the stable vacuum. In traversing this
valley the landscape changes and the typical fluctuations 
will be different from those felt in the metastable regime.
The constants parametrising the 
effective equation of motion must reflect this change. 

Our goal in this Section is to construct \rt{such a ``classical
tunneling''} \at{an effective}
description of the transition to the stable state
by carefully tracing the time evolution of the OP. 
This will be achieved by projecting the microscopic 
equation of motion onto the homogeneous mode and
phenomenologically parametrising it similarly to Eq.(\ref{eqfriction}).

\begin{figure}   
\begin{center}
\includegraphics[width=12cm]{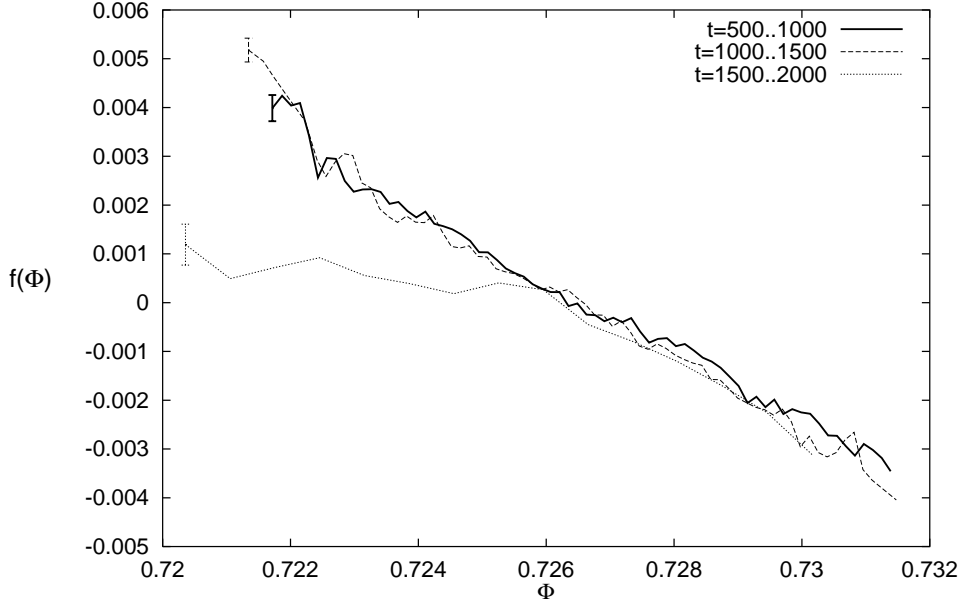}
\end{center}               
\caption{\label{phi} The force $f(\Phi)$ as the function of the order
parameter on $100\times100$ lattice with $h=-0.04/\sqrt{6}$. 
Each curve corresponds to time intervals of length $500|m|^{-1}$.
The interval corresponding to the continous curve ends $1000|m|^{-1}$
before the transition, the long dashed curve refers to the next
time interval and the short dashed curve is the force measured during the last 
time interval preceding the transition. One clearly observes the bending
down left from the central OP-value in the metastable regime. Typical
error bars are shown on the leftmost points of the three curves.}
\end{figure}

As long as the system is far from the narrow valley of the instability
the force is time independent and agrees with the force derived from the
perturbative effective potential according to the part of Fig.~\ref{fphi}
corresponding to $\Phi>0.9$. When \rt{the external magnetic field brings}
the system \at{arrives} close\rt{r} to the entrance of the unstable valley,
$(\Phi\approx0.8)$, 
the soft modes start to be important. This is reflected in the slight
glitch in the leftmost piece of the measured force law in Fig.~\ref{fphi}.
A sequence of glitches results in a situation depicted in Fig.~\ref{phi}. 
It shows the force acting on the OP
in three successive time intervals preceding the event of tunneling for 
$h=-0.04/\sqrt{6}$. The $f(\Phi )$ curves determined using
Eq.~(\ref{eqfriction}) in the disjoint time 
subintervals coincide within error bars for all, but the last one. 
In this last interval preceding directly the transition towards the
direction of negative $\Phi$ values,
the fitted force bends down and its average becomes (a small positive) 
constant. This is characteristic feature of the instant when the system finds 
the entrance
into the unstable potential valley. The \rt{flattening of the potential, the}
vanishing of the force is an indication for the dynamical realization 
of the Maxwell-cut. The OP moves fast through the valley and the method
of fitting the trajectory to Eq.~(\ref{eqfriction}) for finding the force
fails due to the insufficient statistics.

The fluctuation moments depicted in Fig. \ref{time_hist}
tell a bit more about this region. The increased values of the moments, 
the renormalized coupling constants in Wilsonian sense 
at vanishing momentum, indicate
the enhanced importance of the soft interactions as the
OP tunnels through the mean field potential barrier. This softening
makes the OP fluctuating with larger amplitude. The valley of
instability is in a surprising manner flatter than the typical landscape
around equilibrium. 
The flatness along the motion of the OP (the mode ${\bf k}=0$ in momentum 
space) comes from the Maxwell-cut.
 The average curvature
of the potential in the transverse directions, i.e. for modes with 
$|{\bf k}|\not=0$, can be estimated by the second functional derivative of the 
two dimensional effective action with cutoff $|{\bf k}|$, which 
can be taken as ${\bf k}^2+V''_{\bf k}(\Phi)$.
The increased MS fluctuation of the OP corresponds to a decrease in
$V''_{{\bf k}=0}(\Phi)$ in the valley. It pushes down
$V''_{\bf k}(\Phi)$ also for small non-vanishing $|{\bf k}|$, in the
low momentum regime which is expected to be the most influenced by the 
changing landscape.

The lesson of Fig. \ref{phi} is that the potential itself should be considered
as a fluctuating quantity. This helps to translate into more quantitative 
terms the above qualitative points. We discuss the projection  of the
microscopic equation of motion \eq{latteq} onto the
zero momentum sector,
\bea
0&=&\ddot\Phi-\Phi+\Phi^3+3\overline{\varphi^2}^V\Phi
+\overline{\varphi^3}^V-h\nonu
&\equiv&\ddot\Phi+V_{\textrm{inst}}'(\Phi),
\label{eq_OP}
\eea
where the symbol $\overline{\varphi^n}^V$ means the space average of 
$\varphi^n$, ($\overline{\varphi}^V=0, \Phi (t,{\bf x})=\Phi (t)+
\varphi ({\bf x},t)$). The instant potential introduced in the second line 
contains a {\it deterministic} piece, which is the sum of the tree-level 
potential
and of the slowly varying part the second and third moments
(eg. $(\overline{\varphi^n}^V)_{det}$). This last 
feature clearly appears graphycally in Fig.\ref{time_hist} and a simple
model based on a two-phase model of the transition period will be constructed 
below to account for it. The remaining oscillating pieces of the moments 
provide the probabilistic fluctuating contribution to the instant potential:
\be\label{inst}
V_{\textrm{inst}}(\Phi)=-h\Phi-\hf\Phi^2+{1\over4}\Phi^4
+(\overline{\varphi^3}^V)_{det}\Phi
+{3\over2}(\overline{\varphi^2}^V)_{det}\Phi^2+\zeta_0\Phi+{3\over 2}
\zeta_1\Phi^2.
\ee
The additive ($\zeta_0$) and the multiplicative ($\zeta_1$) noises are given 
by the differences
\bea
\zeta_0(t)&=&\overline{\varphi^3}^V(t)-(\overline{\varphi^3}^V)_{det},\nonu
\zeta_1(t)&=&\overline{\varphi^2}^V(t)-(\overline{\varphi^2}^V)_{det}.
\label{zetadef}
\eea
\at{Note that no friction terms can be introduced in a natural way into
the system \hbox{(\ref{eq_OP}-\ref{zetadef})}. Therefore we face the intriguing
question, how irreversibility is realised in such a system. The
time-correlation matrix $\overline{\zeta_i(t)\zeta_j(t+\tau)}^t$
appears to us to be the key object for its investigation, to which we
plan to return in the future.}
\rt{
Note that the breakdown of the time reversal invariance is expected to result
from the cross-correlation between the two noises rather than the effective 
equation of motion \eq{eqfriction} with a single noise and a friction term.}

In order to gain more insight how this works we build into 
Eq.\eq{eq_OP} the consequences of the mixed two-phase picture 
of the phase transformation. The microscopical basis for this picture
is provided by the thermal nucleation whose quantitative discussion is
given for completeness in the Appendix.

More specifically, we assume that the space can be splitted into sharp 
domains (neglecting the thickness of the walls in between), where the field is 
the sum of the constant background values 
$\Phi_{0\pm}$ and the fluctuations $\tilde\varphi_{\pm}$ around it,
\be
\Phi_\pm({\bf x},t)=\Phi_{0\pm}+\tilde\varphi_\pm({\bf x},t).
\ee
We assume local equilibrium in both phases, based on the smooth evolution
of the temperature as displayed in Fig.\ref{k_hist}.

The actual value of the order parameter is determined by the surface
ratio $p(t)$
occupied by the stable phase:
\be
{\Phi(t)}=p(t)\Phi_{0-}+(1-p(t))\Phi_{0+}.
\ee
Simple calculation then yields
\bea
(\overline{\varphi^2}^V)_{det}(t)&=&
\frac{\Phi_{0+}-\Phi(t)}{\Phi_{0+}-\Phi_{0-}}
\left(\overline{\Phi_-^2({\bf x},t)}^V-\overline{\Phi_+^2({\bf x},t)}^V\right)
+\Phi_{0+}^2
-\Phi^2(t)+\overline{{\tilde\varphi_{+}}^2}^V,\nonumber\\
(\overline{\varphi^3}^V)_{det}(t)&=&
\frac{\Phi_{0+}-\Phi(t)}{\Phi_{0+}-\Phi_{0-}}
\left(\overline{\Phi_-^3({\bf x},t)}^V-\overline{\Phi_+^3({\bf x},t)}^V\right)+
\overline{\Phi_+^3({\bf x},t)}^V-3\Phi(t)(\overline{\varphi^2}^V)_{det}(t)-
\Phi^3(t),
\label{bubble_moments}
\eea
where the volume averages should be read off the corresponding 
equilibria on the two sides of the transition. If one takes the values of 
$\Phi_{0\pm},\overline{\tilde \varphi^n_{\pm}}^V$ 
from the respective equilibria determined in the same 
simulation, a quite accurate description of the shape of the
two fluctuation moments arises in the whole transition region and its close 
neighbourhood \at{using $\Phi(t)$ to parametrize their $t$-dependence. 
(Note that $\overline{\Phi^n_\pm(x,t)}^V$ does not depend on time.)
The deterministic part of the moments are therefore well-defined functions
of $\Phi$.} In this way, a simple explicit construction can be given for the
effective noisy equation of the OP-motion, if the above pieces are 
supplemented by the correlation characteristics of the two kinds of noises.

The RMS (root mean square) fluctuations found numerically on 
the transition
part of the trajectory and its close neighbourhood are the following:
\be
\sqrt{\overline{\zeta_0^2(t)}^t}=0.0063(10),\quad \sqrt{\overline
{\zeta_1^2(t)}^t}=0.0054(10), \qquad \textrm{for}\, N=64
\ee
(its magnitude increases with the size of the system).
The magnitude of their equilibrium cross correlation was found $\sim 10^{-5}$.

As a corollary of this construction one can demonstrate the absence of the
deterministic part of the acceleration of the order parameter in the 
transition period. 

Substituting 
Eq.(\ref{bubble_moments}) into Eq.(\ref{eq_OP}), one finds for the 
deterministic part of the force,
\be
f\left(\Phi\right)=\left(-1+\frac{\vavr{\Phi_+^3({\bf x},t)}-
\vavr{\Phi_-^3({\bf x},t)}}
{\Phi_{0+}-\Phi_{0-}}
\right)\Phi+\frac{\Phi_{0+}\vavr{\Phi_-^3({\bf x},t)}-
\Phi_{0-}\vavr{\Phi_+^3({\bf x},t)}}{\Phi_{0+}-\Phi_{0-}}-h.
\label{phieom}
\ee
The average of the equations of motion in the respective equilibria,
\be
\la\Phi_\pm^3({\bf x},t)\ra-\Phi_{0\pm}-h=0
\ee
implies the vanishing of the deterministic force in
Eq.~(\ref{phieom}), when exploiting the equality of the volume and the
ensemble average in this case.
Eq.~(\ref{eq_OP}) transforms into
\be
\ddot\Phi(t)+\zeta_0(t)+3\zeta_1(t)\Phi(t)=0.
\ee
This is the dynamical realization of the Maxwell-construction holding when
the mixed phase model with local equilibrium is valid.

If the force is calculated from the 
full instant potential, $V(\Phi)_\textrm{inst}$ in Eq. \eq{inst},
it depends parametrically on the 
moments $\overline{\varphi^2}^V(t)$ and $\overline{\varphi^3}^V(t)$. The
approximate trajectory
$\Phi_\textrm{inst}(t)$, defined by minimising $V(\Phi)_\textrm{inst}$
with respect to $\Phi$, where the moments are taken from the
numerically determined time evolution, reproduced accurately
the observed OP, $\Phi(t)$ with the following RMS/unit time:
\be
\sqrt{\overline{(\Phi (t)-\Phi_{\textrm {inst}}(t))^2}^t}=0.0014(5),
\qquad \textrm{for}\, N=64.
\ee
This construction interpretes the OP-trajectory 
as a continous deformation of the instant potential with the OP "sitting"
permanently in the actual minimum. Notice that such motion is possible 
only if the OP continously undergoes some sort of dissipation.

The vanishing of the acceleration was checked by comparing the computer
generated OP trajectory in the transition region with a ballistic
motion in a viscous medium.
In particular it was tested that the ratio $h/\dot\Phi$ 
is nearly time and $h$ independent for small enough $h$. 
The dynamical friction measured by the above ratio tends to a
constant with decreasing $h$ at fixed lattice size.
Although the error for \at{its} value in each individual transition is 
rather small, the central values obtained in different runs fluctuate
quite strongly, which leads eventually to a large error of the mean 
calculated as an average of runs with different initial configurations
having the same energy density.

It is worth to note that the order of magnitude of these ``renormalized'', 
i.e.~IR determined 
values of the friction coefficient agree with the peak value in 
Fig.~\ref{atdep} for $N=64$. The ``running'' friction
coefficient determined by the blocking in time, however, drops as
$b_t$ is increased and takes extremely small values at $b_t\approx20$,
the average time length of the ballistic fits for the transition. 
This serves an example
for the dependence of the renormalization group flow on the details
of the blocking. It remains to be understood whether the agreement
between the peak value and the ballistic fit is an accident
or follows from the internal dynamics.

\section{Conclusions and future directions} 
In this paper we have investigated in detail the decay of the false vacuum in 
a classical lattice field theory based exclusively on the effective theory
of the order parameter. Two versions of the effective theory were
reconstructed from its trajectory derived from the microscopical equations
of the theory. The first refers to the (meta)stable branch of the
motion. The second one which takes into account the existence of a mixed phase
during the transition period describes very well the transition
together with its neighbourhood. The first equation has the form of 
conventional mechanical motion taking place in a dissipative noisy 
environment. The dissipation, the dynamical breakdown of the time
inversion symmetry was found only for times
longer then the minimal microscopic time scale of the system, 
the autocorrelation time of the noise.
\rt{The dissipative nature of the second equation is \at{probably}
realised, in the absence of any force term containing odd
order time derivatives, by the presence of two cross-correlated noise terms.}

As a corollary we find that the effective phenomenological OP-theory for 
the decay of the false vacuum through a narrow but flat valley
demonstrates the presence of a dynamical Maxwell-construction.

\at{The next possible directions of extending this work include the
investigation of the effect of the quantum fluctuations on the large
amplitude thermalization of the one-component $\Phi^4$ theory
\cite{Boyan95,Khleb96}.
The effective equation of motion for the order parameter of a quantum
field theory can be studied by breaking up the field $\hat\Phi({\bf x},t)$
into a ``classical'' and a quantum part:
\be
\hat\Phi({\bf x},t)=\Phi_{\textrm{cl}}({\bf x},t)+\hat\phi({\bf x},t),
\qquad
\langle\hat\Phi({\bf x},t)\rangle=\Phi_{\textrm{cl}}({\bf x},t).
\ee
The quantum effects enter the equation of $\Phi_{\textrm{cl}}({\bf
x},t)$ through $\langle\hat\phi^n({\bf x},t)\rangle$, $n=2,3$
analogously to Eq.~(\ref{eq_OP}). The method of the mode-function
expansion \cite{Boyan95,Khleb96,baacke98}
displays explicitly the way the quantum effects
$(\sim\hbar)$ enter the ``classical'' equations of motion. In this
method one solves the time evolution of the coefficient functions of
the expansion
\be
\hat\phi({\bf x},t)=\sum_{\bf Q}
\left[
f_{\bf Q}({\bf x},t)a_{\bf Q}+
f^*_{\bf Q}({\bf x},t)a^+_{\bf Q}
\right],
\ee
where $[a_{\bf Q},a^+_{\bf Q'}]=\delta_{\bf Q Q'}$. The initial
conditions are fixed by requiring the canonical commutation relations
to be fulfilled at $t=0$:
\be
f_{\bf Q}({\bf x},0)=\sqrt{\frac{\hbar}{2\omega_{\bf Q}V}}e^{-i{\bf
Qx}},
\qquad
\dot f_{\bf Q}({\bf x},0)=i\sqrt{\frac{\hbar\omega_{\bf Q}}{2V}}e^{-i{\bf Qx}}.
\ee
The quantum effects relative to Eq.~(\ref{latteq}) show up as ${\cal
O}(\hbar)$ corrections to the coefficients of the effective equation
of motion of $\Phi_{\textrm{cl}}({\bf x},t)$. The adequate form of the
effective equation of motion for $\overline{\Phi_{\textrm{cl}}({\bf
x},t)}^V$ represents a further challenge \cite{workinprogress}.
}

\rt{
The next possible directions of extending this work include
the investigation of the effect of the quantum fluctuations on
the large amplitude thermalisation of the one-component $\Phi^4$ theory
\cite{Boyan95,Khleb96}
and the extension of the results for models with continuous internal symmetry
describing the dynamics of the inflaton field
coupled to the Higgs+Gauge systems \cite{Bellid99,Boyan98}.
}

\at{
The extension of our results to classical field theories with
continous internal symmetry describing the dynamics of the inflaton
field coupled to the Higgs+Gauge system \cite{Bellid99,Boyan98} is also of
actual interest.
}

\section*{Appendix: The nucleation picture}

In this section we analyse the transition using the more conventional
statistical approach of thermal nucleation theory.

From the study (on lattices up to $N=512$) of the detailed microscopical field 
configurations it turned out that the phase transformation starts by the 
nucleation of a single bubble of the stable ($\Phi <0$) phase. 
Late-coming further large bubbles are aggregated to it as
well as the small size (consisting of $n_{bubble}<50-60$ joint sites) 
bubbles. Following the nucleation the expansion rate of the large bubble 
governs the rate of change of the order parameter.
To a very good approximation each individual transition could have been 
characterised by a constant value of $\dot\Phi$ in this interval. Two 
mechanisms are known to lead to this behavior. The first is scattering of hard
waves (``particles'') off the bubble wall, while in the second
the expansion velocity is limited by the diffusive 
aggregation of smaller bubbles \cite{aggreg}. 
The latter process seem to be the dominant in the kinetic Ising model 
\cite{Binder74}.

\begin{figure}
\begin{center}
\includegraphics[width=12cm]{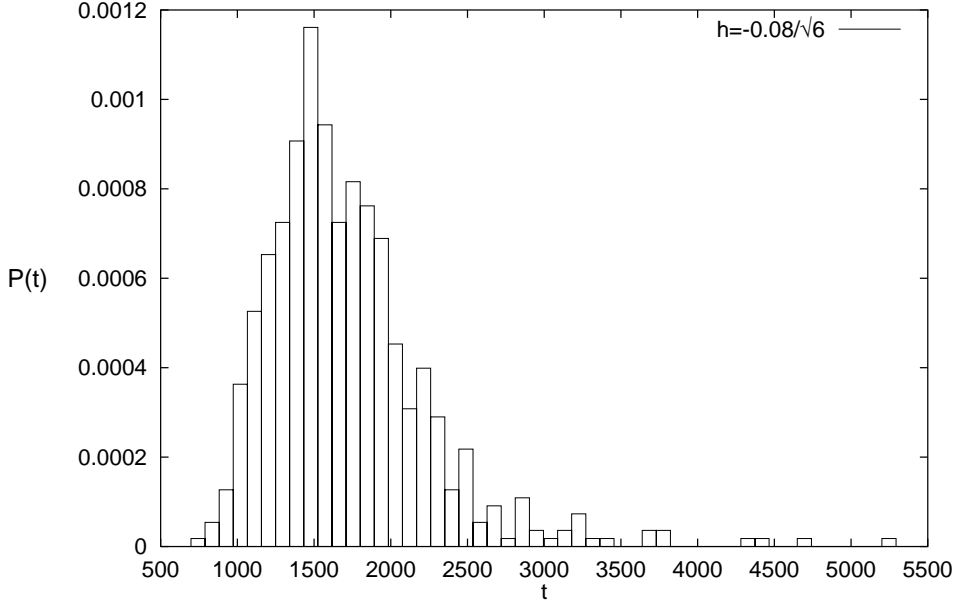}
\end{center}
\caption{\label{release_hist} The normalised
"release" time histogram counting the frequency of a certain time
moment-bin when the order parameter first takes a negative value. The figure
represents the statistics based on 750 events at $h=-0.08/\sqrt{6}$ on an
$N=64$ lattice.}
\end{figure}

The statistics of the ``release'' time of the supercritical bubble from the 
metastable state shows at first sight rather peculiar characteristics.
The binned histogram for larger values of $h$ shows an asymmetric 
peaked structure which apparently deviates from the exponential distribution 
characterising the thermal nucleation scenario (Fig.\ref{release_hist}).
The very early transitions ($t<t_{max}$) seem to be suppressed for small
values of $h$. We expect they correspond to transitions, which happen before
the system reaches the metastable state. 

The fall-off for $t>t_{max}$
starts nearly exponentially, but the histogram develops a very long tail
when $h\rightarrow 0$. The bigger sample is used for estimating 
the probability distribution, the more suppressed is the weight of these 
events in the normalised distribution. In practice this means a longer 
time interval where a good linear fit can be obtained to the 
log-linear histogram. Eventually, the separation of a clean 
exponential signal is possible, and the slope can be compared with 
predictions of the nucleation theory. 

\begin{figure}
\begin{center}
\includegraphics[width=12cm]{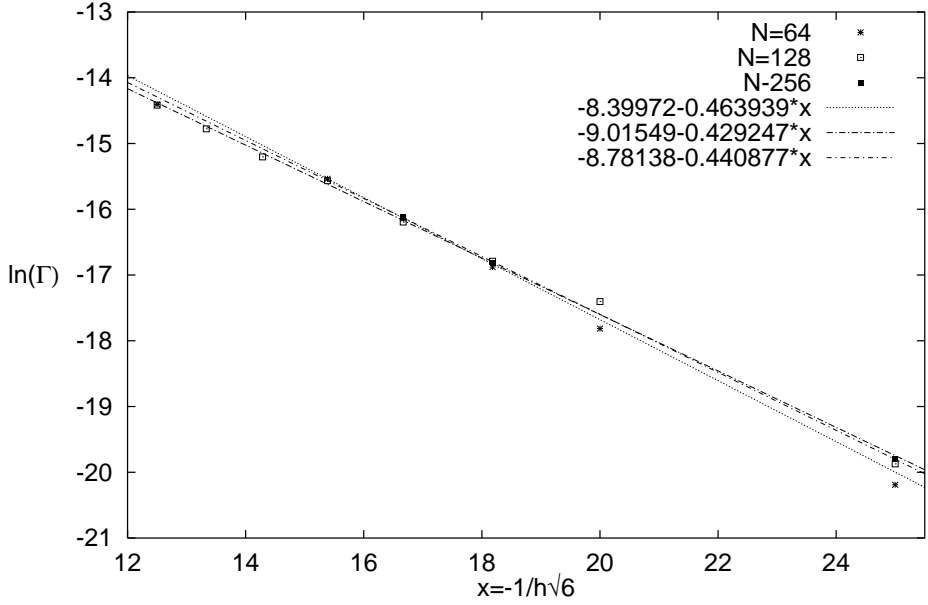}
\end{center}
\caption{\label{t_size}
Nucleation rate for unit lattice surface
versus the reciprocal of the magnetic field for different
lattice sizes. The error bars are smaller than the size of the symbols
representing the data points.}
\end{figure}

The nucleation rate is proportional to the volume of the system $(N^2)$. In
Fig.\ref{t_size} we see evidence for the size independence of the 
transition rate per unit surface. The logarithm of this quantity can be 
estimated following the standard nucleation theory
\cite{Langer67,Coleman80,Gleiser93}:
\be
\ln \Gamma =K-{S_2\over T}.
\ee
$S_2$ is determined approximately by the action of the bounce solution, 
connecting the
two vacua. In this investigation we have studied at fixed temperature 
the $h$-dependence of $S_2/T$, assuming the $h$-independence of $K$. 
In the thin wall approximation the following expression is used
based on the tree level potential:
\be
(S_2)_d=-2\pi\bar r^2h\sqrt{\frac{6}{\lambda}}|m|+8\sqrt2\pi\bar r
\frac{|m|^3}{\lambda},
\ee
(the first term on the right hand side corresponds to the volume energy of
a bubble of radius $\bar r$, 
the second one to its surface energy). For the critical bubble
size a very simple expression is obtained for the exponent of the rate 
in terms of dimensionless quantities:
\be
{S_2(\textrm{thin~wall})\over T}={16\pi\over\sqrt{6}}{|m|^5\over\lambda^{3/2}}
{1\over h_dT_d}={4\pi\over9}{1\over hT}.
\ee

The linear dependence on $1/h$ is fulfilled in our numerical calculations
very well, but the predicted action
is more than one order of magnitude larger than what can be derived from the 
slope of Fig.\ref{t_size}: $S_2/T (\textrm{measured})\approx 0.1/(hT)$. 
If one relaxes the
thin wall approximation and solves numerically the two-dimensional bounce 
equation directly for several $h$, one finds $S_2(\textrm{bounce})/T=0.78/(hT)$.

Further 
improvement can be obtained by applying the temperature corrected effective 
potential in the bounce equation. We have determined the parameters of the
$T$-dependent potential directly from our numerical calculation
 in the following way.
The restoring force was measured for a certain $h$ around both the stable and
the metastable minima as decribed in section 4.
Next an interpolating fit has been constructed of the form: 
$h_{\textrm{eff}}+m^2_{\textrm{eff}}\Phi+\lambda_3\Phi^2+
\lambda_{\textrm{eff}}/6\Phi^3$. It turned out that in the best fit
$\lambda_3\approx 0$ is fulfilled always, while the values of the other
coefficients are not too far from their tree level values.
Then a bounce solution can be built on the corresponding (real!) 
potential which
includes the temperature corrections. The final result is
quite close to the measured value of the rate logarithm:
$S_2(T,\textrm{bounce})/T=0.29/(hT)$. The same real interpolation can be built
on the neighbourhood of the minima of the 2-loop $T$-dependent 
effective potential, leading to
$S_2(T, \textrm{bounce})/T=0.2/(hT)$.
By common experience in the surface tension 
simulations a factor of 2-3 difference in $S_2$ is
expected to arise relative to the mean field theory.

We conclude, that the finite temperature corrections are important for 
quantitative
treatment of the nucleation rate within the thermal nucleation theory.

\begin{figure}                       
\begin{center}                
\includegraphics[width=12cm]{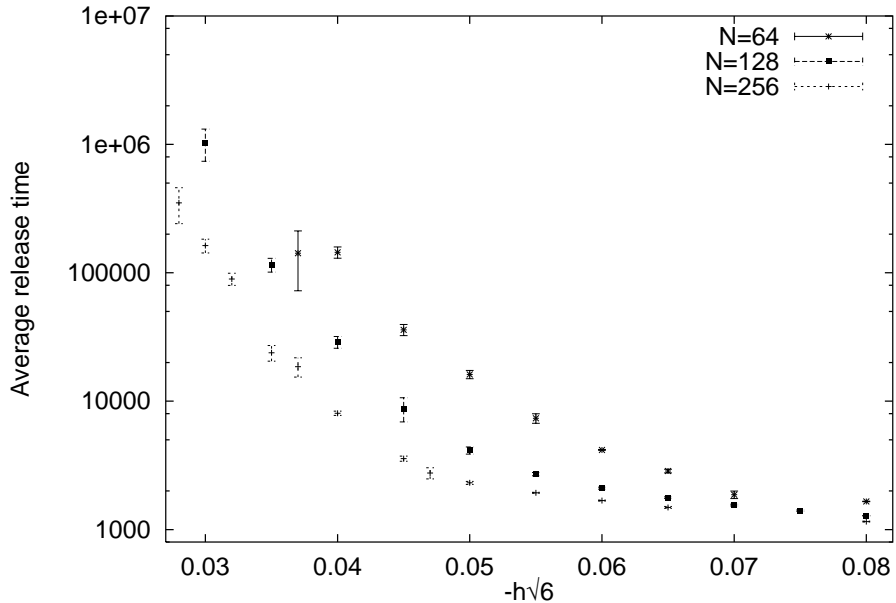}
\end{center} 
\caption{\label{t_release} The average "release" time as a function
of $h$ measured on $N=64,128,256$ lattices.} 
\end{figure}

Next we turn to the discussion of the possible nucleation
threshold at small $h$.
The average ``release time'' increases for fixed lattice size with 
decreasing $h$ (Fig.\ref{t_release}). 
On a lattice of fixed size we found small $h$ values for which we could 
not detect any transition, what makes very probable the existence of a 
threshold
value of the external field $h_{th}(N)$. We did not attempt to locate this
value beyond the simple hyperbolic fits to the few largest 
$\langle t_{release}\rangle$ values. The value of $h_{th}(N)$ remains stable 
when the smallest $h$ is excluded from the fit, therefore we conclude that a 
threshold magnetic field exists similar to the case of the kinetic Ising 
model \cite{Binder74}. The value of $h_{th}$ decreases when the lattice size
is increased. Intuitively we expect $h(\infty )=0$, but with the three
lattices studied by us this conjecture cannot be demonstrated.

\section*{Acknowledgements}
The authors are grateful for valuable discussions to C. Gagne, J. Hajdu,
A. Jakov\'ac, Z. R\'acz
and H.J. de Vega. They acknowledge the use of computing resources
generously provided by the  University of Bielefeld and the Dept.
of Computer Science and the Inst. for Theoretical Physics
of the E{\"o}tv{\"o}s University. This research has been supported by CNRS and
the Hungarian Science Fund (OTKA).

\end{document}